\documentclass[12pt,preprint]{aastex}
\usepackage{amssymb}
\usepackage{color}

\slugcomment{Not to appear in Nonlearned J., 45.}

\shorttitle{The disk evaporation model for low-luminosity active
galactic nuclei} \shortauthors{Erlin Qiao et al.}

\begin{document}



\title{The disk evaporation model for the spectral features
of low-luminosity active galactic nuclei}


\author{Erlin Qiao\altaffilmark{1}, B. F. Liu \altaffilmark{1},
Francesca Panessa \altaffilmark{2} and J. Y. Liu \altaffilmark{3}}

\email{qiaoel@nao.cas.cn}

\altaffiltext{1}{National Astronomical Observatories, Chinese
Academy of Sciences, Beijing 100012, China}

\altaffiltext{2}{INAF - Istituto di Astrofisica e Planetologia
Spaziali di Roma (IAPS), Via del Fosso del Cavaliere 100, 00133
Roma, Italy}

\altaffiltext{3}{National Astronomical Observatories /Yunnan
Observatory, Chinese Academy of Sciences, P.O. Box 110, Kunming
650011, P. R. China}


\begin{abstract}

Observations show that the accretion flows in low-luminosity active
galactic nuclei (LLAGNs) probably  have a two-component structure
with an inner hot, optically thin, advection dominated accretion
flow (ADAF) and an outer truncated cool, optically thick accretion
disk. As shown by Taam et al. (2012), the truncation radius as a
function of mass accretion rate is strongly affected by including
the magnetic field within the framework of disk evaporation model,
i.e., an increase of the magnetic field results in a smaller
truncation radius of the accretion disk. In this work, we calculate
the emergent spectrum of an inner ADAF + an outer truncated
accretion disk around a supermassive black hole based on the
prediction by Taam et al. (2012). It is found  that an increase of
the magnetic field from $\beta=0.8$ to $\beta=0.5$ (with magnetic
pressure $p_{\rm m}=B^2/{8\pi}=(1-\beta)p_{\rm tot}$, $p_{\rm
tot}=p_{\rm gas}+p_{\rm m}$) results in an increase of $\sim 8.7$
times of the luminosity from the truncated accretion disk, meanwhile
results in the peak emission of the truncated accretion disk
shifting towards a higher frequency by a factor of $\sim 5$ times.
We found that the equipartition of gas pressure to magnetic
pressure, i.e., $\beta=0.5$, failed to explain the observed
anti-correlation between $L_{\rm 2-10 keV}/L_{\rm Edd}$ and the
bolometric correction $\kappa_{\rm 2-10 keV}$ (with $\kappa_{\rm
2-10 keV} = L_{\rm bol}/L_{\rm 2-10 keV}$). The emergent spectra for
larger value $\beta=0.8$ or $\beta=0.95$ can well explain the
observed $L_{\rm 2-10 keV}/L_{\rm Edd}$-$\kappa_{\rm 2-10 keV}$
correlation. We argue that in the disk evaporation model, the
electrons in the corona are assumed to be heated only by a transfer
of energy from the ions to electrons via Coulomb collisions, which
is reasonable for the accretion with a lower mass accretion rate.
Coulomb heating is the dominated heating mechanism for the electrons
only if the magnetic field is strongly sub-equipartition, which is
roughly consistent with observations.

\end{abstract}


\keywords{accretion, accretion disks
--- Black hole physics --- galaxies: active --- X-rays: galaxies}

\section{Introduction}

Active galactic nuclei (AGNs) are believed to be powered by
accretion onto super-massive black holes, which are strong sources
emitting from radio to X-rays. The luminous AGNs, mainly including
radio quiet quasars and bright Seyfert galaxies, are believed to be
powered dominantly by a geometrically thin, optically thick,
accretion disk extending down to the innermost stable circular
orbits (ISCO) of a black  hole (Shakura \& Sunyaev 1973; Shields
1978; Malkan \& Sargent 1982; Kishimoto, Antouncci \& Blaes 2005;
Shang et al. 2005; Liu et al. 2012). Observations show that
accretion flows in low-luminosity active galactic nuclei (LLAGNs)
are very different. A two-component structure of the accretion flow
with an inner hot, optically thin, advection dominated accretion
flow (ADAF), an outer cool, optically thick truncated accretion
disk, and a jet are supported by a number of observational evidence
for LLAGNs (Ho 2008 and the references therein). The observational
evidence for such a structure in LLAGNs comes mainly from the
broadband emission from the radio to X-rays (Lasota et al. 1996;
Quataert et al. 1999; Di Matteo et al.2000, 2003; Yuan et al. 2009;
Li et al. 2009). For example, by fitting the spectral energy
distribution (SED) of NGC 1097, Nemmen et al. (2006) found that the
optical and X-ray portion of the SED can be well fitted by an inner
ADAF and an outer accretion disk truncated at $225$ Schwarzschild
radii, and the observed radio emission can be very well interpreted
by the synchrotron emission of a relativistic jet modeled within the
framework of the internal shock scenario (Yuan et al. 2005). The
presence of a jet in LLAGNs has been clearly detected by the VLBI
radio observations (Falcke et al. 2000; Nagar et al. 2001).
Meanwhile, theoretically, because of the positive Bernoulli
parameter of the hot ADAF, it is probably that the formation of the
bipolar jets are driven by the ADAF (Narayan \& Yi 1995b; Blandford
\& Begelman 1999; Meier et al. 2001; Yuan et al. 2012a, b).

The evidence for the presence of an inner ADAF in LLAGNs is inferred
from the very low luminosity and the very lower radiative efficiency
estimated from the available mass supply rate (Ho 2009). In terms of
Eddington luminosity, LLAGNs often have $\lambda \lesssim 0.01$
contrary to the luminous AGNs with $\lambda \gtrsim 0.01$  (with
$\lambda=L_{\rm bol}/L_{\rm Edd}$, $L_{\rm Edd}=1.26 \times 10^{38}
M/M_{\odot} \rm \ erg \ s^{-1}$) (Panessa et al. 2006). The observed
anti-correlation between the hard X-ray index $\Gamma_{\rm 2-10 \rm
keV}$ and $\lambda$ implies that LLAGNs may be dominated by ADAF (Gu
\& Cao 2009; Younes et al. 2011, 2012). Xu (2011) collected a LLAGN
sample composed of 49 sources including 28 local Seyfert galaxies
and 21 low-ionization nuclear emission-line regions (LINERs) with
optical/UV and X-ray observations. It is found that there is a
strong anti-correlation between $\alpha_{\rm ox}$ and Eddington
ratio $\lambda$ for the sources with $\lambda \lesssim 10^{-3}$,
which supports that ADAF model is a very promising candidate in
LLAGNs.

The evidence for the truncation of the accretion disk in LLAGNs is
inferred from the lack of a `big blue bump', instead of a `big red
bump' (Nemmen et al. 2012). The very weak or absent broad iron $\rm
K_{\rm \alpha}$ line, which is attributed to X-ray fluorescence off
of an optically thick accretion disk extending down to a few
Schwarzschild radii of luminous AGNs, also supports that the
accretion disk truncates at a larger radius ($\sim 100-1000$
Schwarzschild radii) off the black hole in LLAGNs (Nandra et al.
2007). Quataert et al. (1999) fitted the optical/UV spectrum of M81
with a mass accretion rate of $0.01$ Eddington accretion rate and an
accretion disk truncated at $100$ Schwarzschild radii. A similar fit
to the optical/UV spectrum of NGC 4579 yielded a higher mass
accretion rate of $0.03$ Eddington accretion rate, while the
truncation radius of the accretion disk is still  $100$
Schwarzschild radii (Quataert et al. 1999). The relationship between
the truncation radius and the mass accretion rate is still
controversial. By fitting the SEDs of a population of LLAGNs, Yuan
\& Narayan (2004) found that the lower Eddington ratio $L_{\rm
bol}/L_{\rm Edd}$ was associated with a larger truncation radius. By
studying 33 PG quasars with Fe $\rm K_{\alpha}$ emission line
detected by the {\em XMM-Newton} survey, Inoue, Terashima \& Ho
(2007) found that the Fe $\rm K_{\alpha}$ line systematically
becomes narrow with decreasing $L_{\rm bol}/L_{\rm Edd}$, which
probably means that the truncation radius of the accretion disk
increases with decreasing mass accretion rate. It is interesting
that the inverse correlation between the truncation radius of the
accretion disk and the mass accretion rate is also found in the
low/hard spectral state of black hole X-ray binaries (Cabanac et al.
2009).

The physical mechanism for determining the truncation of the
accretion disk has been proposed by many authors (Honma 1996;
Manmoto \& Kato 2000; Lu et al. 2004; Spruit \& Deufel 2002;
Dullemond \& Spruit 2005). One of the promising model among them is
the disk evaporation model, which was first proposed by Meyer \&
Meyer-Hofmeister (1994) for dwarf novae, established for black holes
by Meyer et al. (2000a, b) and modified by Liu et al. (2002), where
the decoupling of ions and electrons and Compton cooling effect are
taken into account. The disk evaporation model has been applied to
explain the spectral state transition in stellar-mass black holes
(e.g., Meyer-Hofmeister \& Meyer 2005; Qiao \& Liu 2009; Qian et al.
2007). The spectral features of an inner ADAF and an outer truncated
accretion disk predicted by the disk evaporation model in
stellar-mass black holes were investigated by Qiao \& Liu (2010,
2012). The first application of disk evaporation model to interpret
the truncation of the accretion disk in LLAGNs was from Liu et al.
(1999), in which a theoretical relation between the truncation
radius and the mass accretion rate was given. Taam et al. (2012)
generalized the disk evaporation model in black hole X-ray binaries
by including the effect of a magnetic field in accretion disk to
apply the model to the truncation of the accretion disk in a large
number of LLAGNs. It is found that the truncation radius of the
accretion disk can be very strongly affected by the magnetic
parameters $\beta$, which is defined as $p_{\rm
m}=B^2/{8\pi}=(1-\beta)p_{\rm tot}$, (where $p_{\rm tot}=p_{\rm
gas}+p_{\rm m}$, $p_{\rm gas}$ is gas pressure and $p_{\rm m}$ is
magnetic pressure), describing the strength of the magnetic field in
the accretion flows. Taam et al. (2012) found that the inclusion of
magnetic field results in a smaller truncation radius compared to
the case without magnetic field.

In this work, based on the prediction by Taam et al. (2012) for the
truncation of the accretion disk, the emergent spectrum of a
two-component structure with an inner ADAF and an outer truncated
accretion disk around a supermassive black hole is calculated. It is
found that the disk evaporation model can roughly reproduce the
observed anti-correlation between the hard X-ray index $\Gamma_{\rm
2-10keV}$ and the Eddington ratio $\lambda$ in LLAGNs (e.g., Gu \&
Cao 2009). As shown by Taam et al. (2012), the truncation radius of
the accretion disk is very sensitive to $\beta$, which consequently
affects the emergent spectrum. Our calculations show that the
equipartition of gas pressure to magnetic pressure, i.e.,
$\beta=0.5$, failed to explain the observed anti-correlation between
$L_{\rm 2-10 keV}/L_{\rm Edd}$ and the bolometric correction
$\kappa_{\rm 2-10 keV}$ (with $\kappa_{\rm 2-10 keV} = L_{\rm
bol}/L_{\rm 2-10 keV}$). The emergent spectra for larger value
$\beta=0.8$ or $\beta=0.95$ can well explain the observed $L_{\rm
2-10 keV}/L_{\rm Edd}$-$\kappa_{\rm 2-10 keV}$ correlation. It is
argued that the sub-equipartition of the magnetic field is
reasonable in the case with a low mass accretion. In section 2, we
briefly introduce the disk evaporation model. The emergent spectra
of disk evaporation model are presented in section 3. Some
applications of the disk evaporation model to LLAGNs are presented
in section 4. Section 5 is the conclusion.

\section{The Model}
\subsection{Truncation of the accretion disk}

We consider a hot corona above a geometrically thin standard disk
around a central black hole. In the corona, viscous dissipation
leads to ion heating, which is partially transferred to the
electrons by means of Coulomb collisions. This energy is then
conducted down into lower, cooler, and denser corona. If the density
in this layer is sufficiently high, the conductive flux is radiated
away. If the density is too low to efficiently radiate energy, cool
matter is heated up and evaporation into the corona takes place. The
mass evaporation goes on until an equilibrium density is
established. The gas evaporating into the corona still retains
angular momentum and with the role of viscosity will differentially
rotate around the central object. By friction the gas loses angular
momentum and drifts inward, and thus continuously drains mass from
the corona towards the central object. This is compensated by a
steady mass evaporation flow from the underlying disk. The process
is driven by the gravitational potential energy released by friction
in the form of heat in the corona. Therefore, mass accretes to the
central object partially through the corona (evaporated part) and
partially through the disk (the left part of the supplying mass).
Such a model for black holes was established by Meyer et al. (2000a,
b). The inflow and outflow of mass, energy, and angular momentum
between neighboring zones were included by Meyer-Hofmeister \& Meyer
(2003). The effect of the viscosity parameter $\alpha$ is
investigated by (Qiao \& Liu 2009), and the effect of the magnetic
parameter $\beta$ is studied by (Qian et al. 2007). The model we
used here is based on Liu \& Taam (2009), in which the structure of
the corona and the evaporation features are determined by the
equation of state, equation of continuity, and equations of momentum
and energy. The calculations show that the evaporation rate
increases with decreasing distance in the outer region of the
accretion disk, reaching a maximum value and then dropping towards
the central black hole. Taam et al. (2012) generalized the results
of disk evaporation model for black hole X-ray binaries by including
the effect of viscosity parameters $\alpha$ and magnetic parameter
$\beta$ in accretion disks around a supermassive black hole. The
maximum evaporation rate and the corresponding radius from the black
hole as functions of $\alpha$ and $\beta$ are given as (Taam et al.
2012),
\begin{eqnarray}\label{dotmax}
\dot m_{\rm max} \approx 0.38\alpha^{2.34} \beta^{-0.41},
\end{eqnarray}
\begin{eqnarray}\label{rmax}
r_{\rm min} \approx 18.80 \alpha^{-2.00} \beta^{4.97},
\end{eqnarray}
where the evaporation rate is in units of Eddington accretion rate
$\dot M_{\rm Edd}$ ($\dot M_{\rm Edd}$ = $1.39 \times 10^{18}
M/M_{\rm \odot} \rm \ g\ s^{-1}$), and the radius is in units of
Schwarzschild radius $R_{\rm S}$ ($R_{\rm S}$= $2.95 \times 10^5 $
$M/M_{\rm \odot} \rm \ cm$).

If the accretion rate is higher than the maximum evaporation rate,
the evaporation can only make a fraction of the disk accretion flow
go to the corona and the optically thick disk is never completely
truncated by evaporation. However, if the mass supply rate is less
than the maximum evaporation rate, the matter in the inner region of
the disk will be fully evaporated to form a geometrically thick,
optically thin accretion inner region, which is generally called
advection-dominated accretion flow (ADAF). The truncation radius of
the disk can be generalized as (Taam et al. 2012),
\begin{eqnarray}\label{tr}
r_{\rm tr} \approx 17.3 \dot m^{-0.886} \alpha^{0.07} \beta^{4.61},
\end{eqnarray}
where $\dot m$ is the mass supply rate from the most outer region of
the accretion disk. If $\alpha$, $\beta$ and $\dot m$ are specified,
we can self-consistently get a two-component structure of the
accretion flow with an inner ADAF and an outer truncated accretion
disk. It can be seen that from Equation (\ref{tr}), the truncation
radius is very weakly dependent on $\alpha$, but strongly dependent
on $\beta$. It is argued that the effect of magnetic fields on
evaporation rate is a competition between its tendency to increase
the evaporation as a result of the energy balance and to decrease
the evaporation as a result of the pressure balance. The additional
pressure contributed by the magnetic fields results in a greater
heating via the shear stress. This effect is similar to an increase
viscosity parameter and leads to an increase of evaporation rate in
the inner region with little effect in the outer region. In this
paper, because we focus on the truncation of the accretion disk in
the outer region, the additional pressure contribution suppresses
the evaporation at all distances as a result of force balance.  The
effect of the magnetic field results in little change in the value
of the maximal evaporation rate, but make the evaporate curve move
systematically inward, which means the truncation radius will
decrease for a given mass accretion rate. Meanwhile, we address that
Equation (\ref{tr}) is a good fit only when the mass accretion rate
is less than half of the maximum evaporation rate, i.e., $\dot m
\lesssim $ $(1/2) \dot m_{\rm max}$=
$0.19\alpha^{2.34}\beta^{-0.41}$$\approx 0.01$ (assuming a standard
viscosity parameter $\alpha=0.3$). When the mass accretion rate is
close to the maximum evaporation rate, the truncation radius
deviates from the power-law expression of Equation (\ref{tr}). In
this case, the truncation radius will be determined by detailed
numerical calculations (Taam et al. 2012).

\subsection{The ADAF model}
Inside the truncation radius, the accretion flows are in the form of
advection dominated accretion flow (ADAF) or radiatively inefficient
accretion flow (RIAF) from our disk-evaporation model (Rees et al.
1982; Narayan \& Yi 1994; Narayan et al. 1998; Quataert 2001;
Narayan \& McClintock 2008, and the references therein). The
self-similar solution of ADAF was first proposed by Narayan \& Yi
(1994, 1995b), with which the spectrum of transient sources A0620-00
and V404 Cyg with lower luminosity were well fitted (Narayan et al.
1996).  Later, the global solution of ADAF are conducted by several
authors (Narayan et al. 1997; Manmoto 1997, 2000; Yuan et al. 1999,
2000; Zhang et al. 2010). The applications of ADAF to interpret the
broadband spectrum of supermassive black holes can be seen, e.g., in
Yuan et al. (2003) for the Galactic center, and by Quataert et al.
(1999) for the LLAGN M81 and NGC 4579 and so on. All of them show
that the self-similar solution is a good approximation at a radius
far enough from the ISCO. For simplicity, in this paper, the
self-similar solution of ADAF is adopted (Narayan \& Yi 1995a, b).
The structure of an ADAF surrounding a black hole with mass $M$ can
be calculated if the parameters including $\dot m$, $\alpha$ and
$\beta$ are specified.

\section{Numerical result}


We calculate the emergent spectra predicted by the disk evaporation
model for a two-component structure composted of an inner ADAF and a
truncated accretion disk around a supermassive black hole with mass
$M$ when the parameters including $\dot m$, $\alpha$ and $\beta$ are
specified. In the calculation, we fix central black hole mass at $
M=10^8 M_{\odot}$, assuming a viscosity parameter of $\alpha=0.3$,
as adopted by Taam et al. (2012) for the spectral fits to LLAGNs.

\begin{figure*}
\includegraphics[width=85mm,height=70mm,angle=0.0]{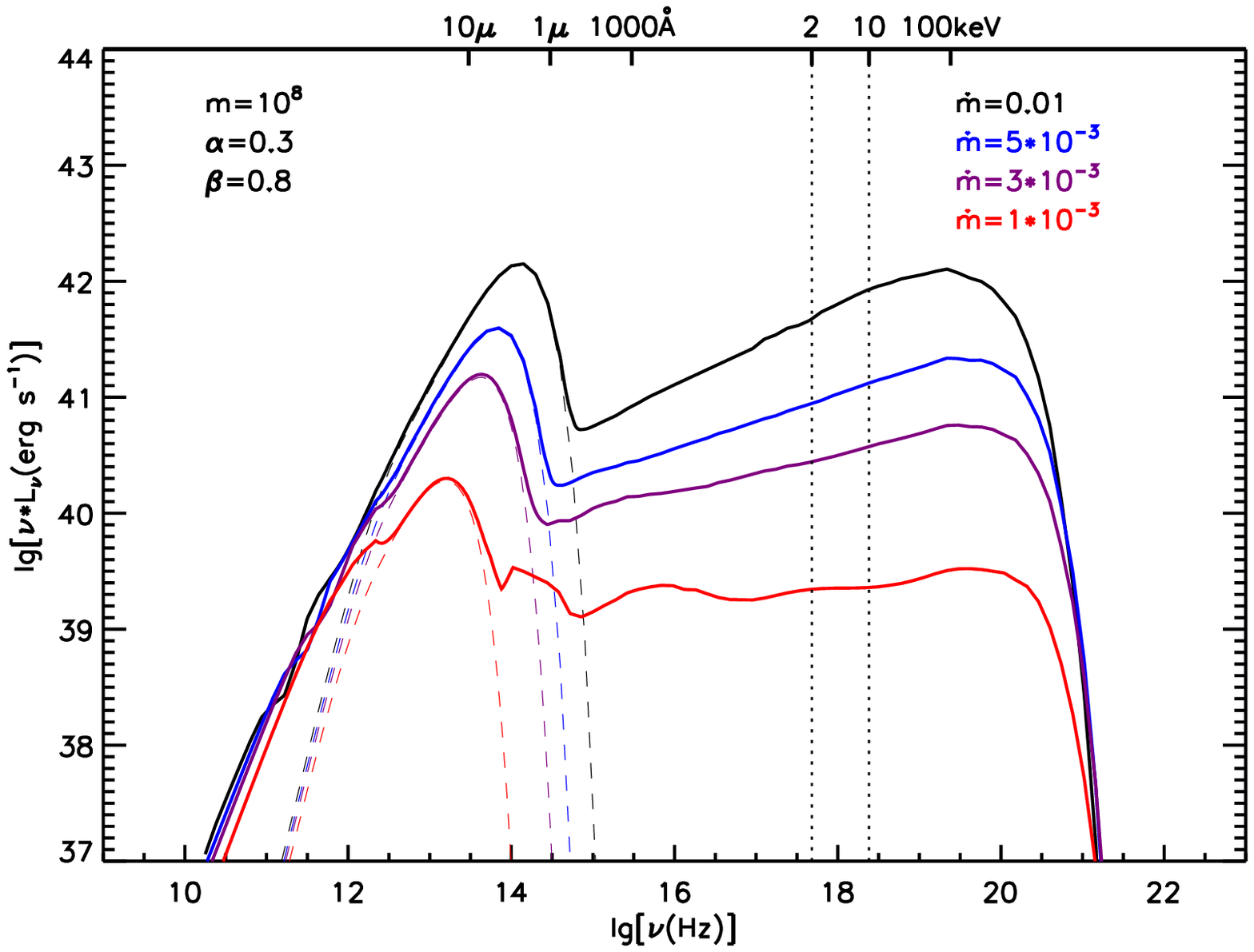}
\includegraphics[width=85mm,height=70mm,angle=0.0]{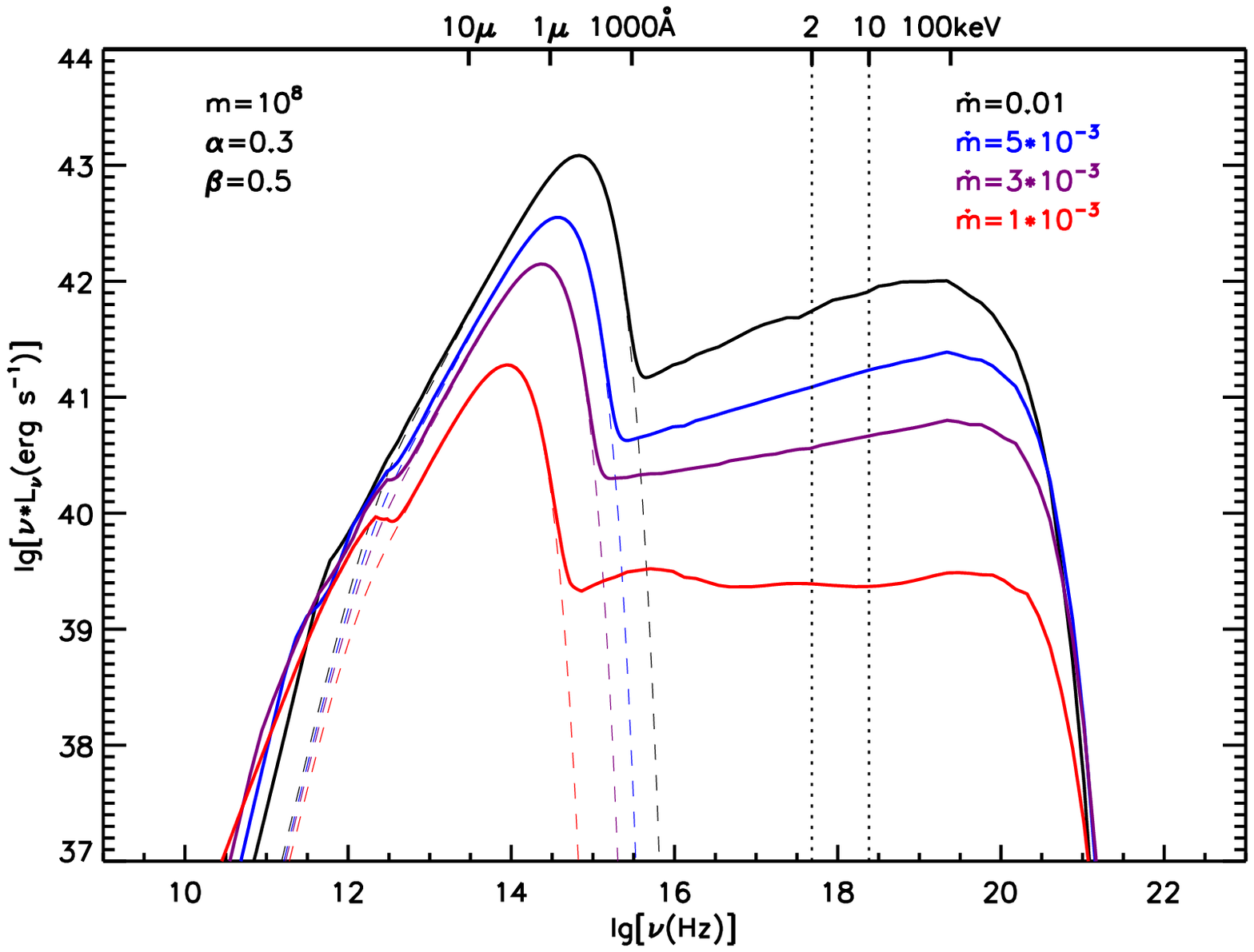}
\caption{\label{108}} Emergent spectra of an inner ADAF and an outer
truncated accretion disk around a black hole predicted by the disk
evaporation model with $M=10^{8} M_{\odot}$ assuming $\alpha=0.3$.
Left panel: $\beta=0.8$ is adopted. From bottom to top, the solid
lines are the combined emergent spectra from an inner ADAF plus an
outer truncated accretion disk for  $\dot m= 10^{-3}$, $3\times
10^{-3}$, $5\times 10^{-3}$ and $0.01$ respectively. The dashed line
are the emergent spectra from the truncated accretion disk. Right
panel: $\beta=0.8$ is adopted, and the meaning of the line style is
same with the left panel.
\end{figure*}

The emergent spectra with mass accretion rates for $\beta=0.8$ are
plotted in the left panel of Figure \ref{108} for mass accretion
rates $\dot m=0.01$, $0.005$, $0.003$, $0.001$. The black-solid line
is the total emergent spectrum for $\dot m=0.01$, and the
black-dashed line is the emission from the accretion disk with a
truncation radius $r_{\rm tr}=310$ predicted by Equation (\ref{tr}).
The maximum effective temperature of the truncated accretion disk is
$\sim 2599 \rm K$, which has a typical emission peaking at $\sim 1.1
\mu m$ (in $L_{\nu}$ vs. $\nu$). The hard X-ray emission between
2-10 keV for $\dot m=0.01$ can be described by a power law with a
hard X-ray index $\Gamma=1.65$, which is produced by the
self-Compton scattering of the synchrotron and bremsstrahlung
photons of the ADAF itself. With a decrease of the mass accretion
rate to $\dot m=0.005$, the predicted truncation radius of the
accretion disk is $r_{\rm tr}=573$ , which has a maximum effective
temperature $\sim 1388 \rm K$ with a peak emission at $\sim 2.1 \mu
m$. Meanwhile, the hard X-ray index between 2-10 keV is
$\Gamma=1.75$. The blue-solid line in the left panel of Figure
\ref{108} is the total emergent spectrum for $\dot m=0.005$, and the
blue-dashed line is the emission from the truncated accretion disk.
With a further decrease of the mass accretion rates to $\dot
m=0.003$ and $\dot m=0.001$, the truncation radii of the accretion
disk are $r_{\rm tr} = 901$ and $r_{\rm tr}=2386$, which correspond
to an effective temperature of the accretion disk of $\sim 874 \rm
K$ and $\sim 321 \rm K$  with peak emissions at $\sim 3.3 \mu m$ and
$\sim 9 \mu m$ respectively. Meanwhile, the hard X-ray indices are
$\Gamma=1.81$ for $\dot m=0.003$ and $\Gamma=1.98$ for $\dot
m=0.001$ respectively. The purple-solid line and the red-solid line
in Figure \ref{108} are the total emergent spectra for $\dot
m=0.003$ and $\dot m=0.001$, and the purple-dashed and the
red-dashed lines are the emissions from the truncated accretion
disks respectively.

\begin{figure*}
\includegraphics[width=85mm,height=70mm,angle=0.0]{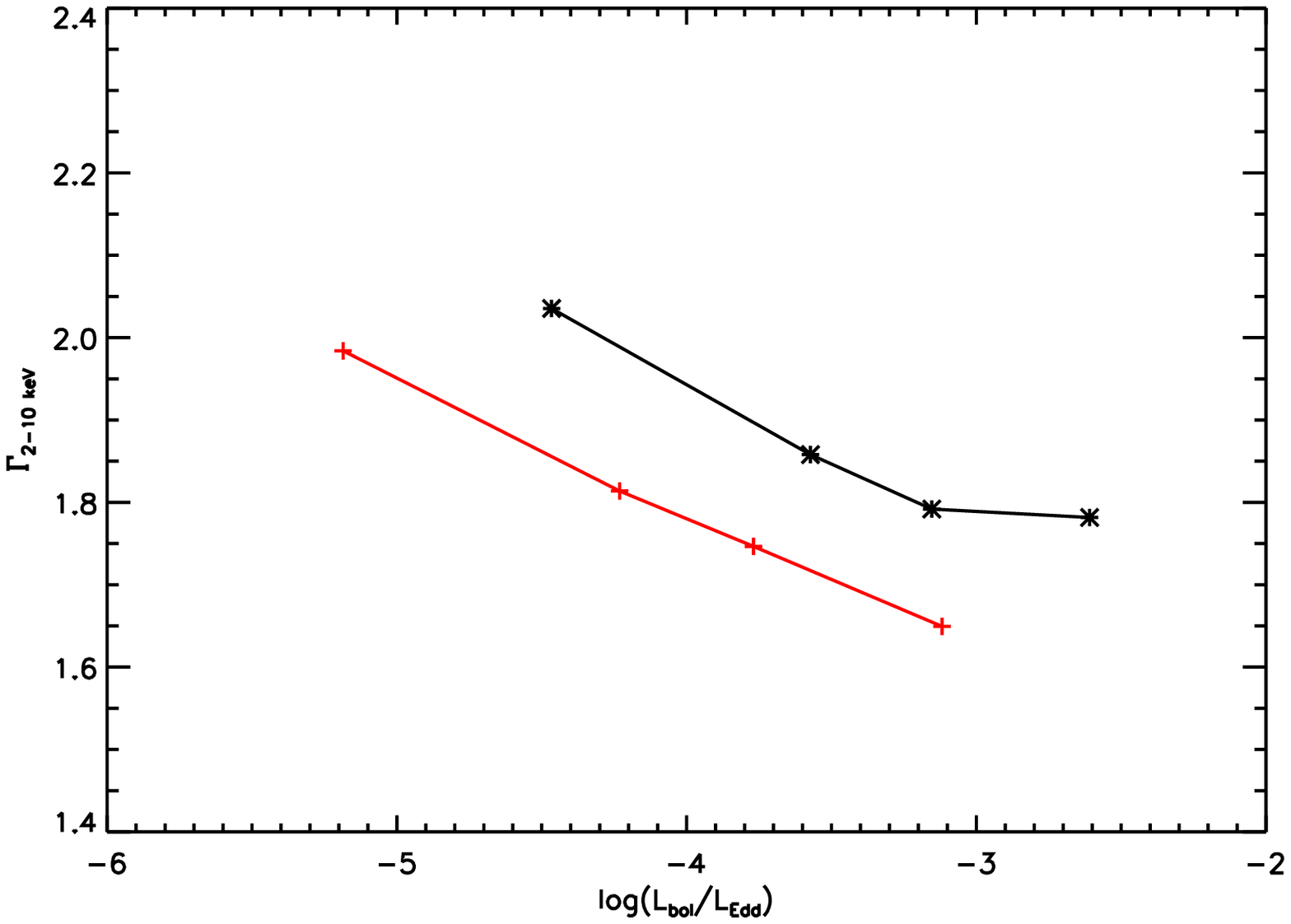}
\caption{\label{gama}} Hard X-ray index $\Gamma_{\rm 2-10keV}$ as a
function of Eddington ratio $L_{\rm bol}/L_{\rm Edd}$. In the
calculations, we fix the black hole mass at $ M=10^8 M_{\odot}$,
assuming a viscosity parameter $\alpha=0.3$. The red line is for
$\beta=0.8$ and the black line is for $\beta=0.5$.
\end{figure*}

We plot the hard X-ray index $\Gamma_{\rm 2-10 keV}$ as a function
of Eddington ratio $L_{\rm bol}/L_{\rm Edd}$ for $\beta=0.8$ in
Figure \ref{gama} with a red line. Here the bolometric luminosity
$L_{\rm bol}$ is calculated by integrating the emergent spectrum. It
is found that there is an anti-correlation between $\Gamma_{\rm 2-10
keV}$ and $L_{\rm bol}/L_{\rm Edd}$, which is qualitatively
consistent with the observations in LLAGNs (Constantin et al. 2009;
Wu \& Gu 2009; Younes et al. 2011, 2012; Xu 2011). This is because,
with the decrease of the mass accretion rate, the electron
temperature of ADAF $T_{\rm e}$ changes only slightly, and it is
always around $10^9 \rm K$ (Mehadevan 1997). However, the decrease
of the mass accretion rate will result in a direct decrease of the
Compton scattering optical depth $\tau_{\rm es}$. So the Compton
parameter $y=4kT_{\rm e}/{m_{\rm e}c^2}$ Max($\tau_{\rm es}$,
$\tau_{\rm es}^2$) of the ADAF decreases with decreasing mass
accretion rate, consequently resulting in a softer spectrum, as also
discovered in the low/hard spectral state of black hole X-ray
binaries (Qiao \& Liu 2010; 2013; Wu \& Cao 2008; Yuan et al. 2005).

The emergent spectra with mass accretion rates for $\beta=0.5$ are
plotted in the right panel of Figure \ref{108} for mass accretion
rates $\dot m=0.01$, $0.005$, $0.003$, $0.001$. The black-solid line
is the emergent spectrum for $\dot m=0.01$, and the black-dashed
line is the emission from the accretion disk with a truncation
radius $r_{\rm tr}=30$ predicted by Equation \ref{tr}. The maximum
effective temperature of the truncated accretion disk is $\sim 1.4
\times 10^4 \rm K$, which has a typical UV emission peaking at $\sim
2075 \rm \AA$.  The hard X-ray emission for $\dot m=0.01$ in 2-10
keV can be described by a power law with a hard X-ray index
$\Gamma=1.78$. For $\dot m= 0.005, 0.003, 0.001$, the truncation
radii of the accretion disk are $r_{\rm tr} = 55.4, 87, 231$
respectively. The maximum effective temperature of the truncated
accretion disk is $\sim 7635 \rm K, \sim 4862 \rm K$  and $\sim 1814
\rm K$ respectively, with the emission peaking at $3798 \rm \AA$,
$5964 \rm \AA$ and  $1.6 \rm \mu m$ respectively. The hard X-ray
indices between 2-10 keV are $\Gamma_{2-10 \rm keV} = 1.79, 1.85,
2.04$ respectively. We plot $\Gamma_{\rm 2-10 keV}$ as a function of
$L_{\rm bol}/L_{\rm Edd}$ for $\beta=0.5$ in Figure \ref{gama} with
the black line. It is also clear that there is an anti-correlation
between $\Gamma_{\rm 2-10 keV}$ and $L_{\rm bol}/L_{\rm Edd}$.

\begin{figure*}
\includegraphics[width=85mm,height=70mm,angle=0.0]{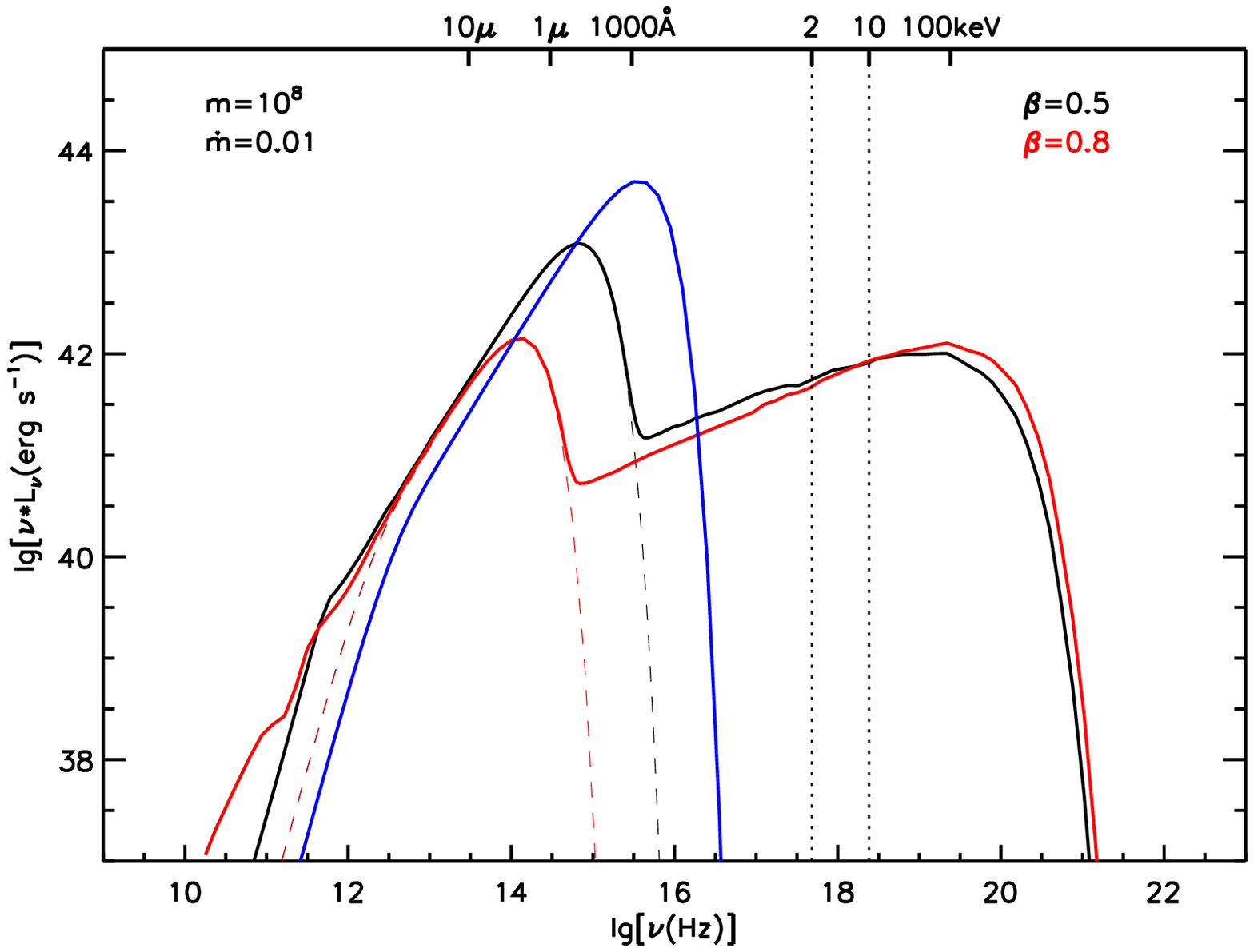}
\caption{\label{disk-beta}} Emergent spectra of an inner ADAF and an
outer truncated accretion disk around a black hole predicted by the
disk evaporation model with $M=10^{8} M_{\odot}$ assuming
$\alpha=0.3$. The solid-black line is for $\beta=0.5$ and $\dot
m=0.01$, where the disk is truncated at $30 R_{\rm S}$. The
red-solid line is for $\beta=0.8$ and $\dot m=0.01$, where the disk
is truncated at $310 R_{\rm S}$. The dashed lines are the emergent
spectra from the truncated accretion disk. The blue line is the
emergent spectrum for $\dot m=0.01$ with the standard accretion disk
extending down to the ISCO of a non-rotating black hole.
\end{figure*}

In order to clearly show the effect of the magnetic parameter
$\beta$ on the spectra, we fix the central black hole mass at $
M=10^8 M_{\odot}$, $\alpha=0.3$ and $\dot m=0.01$ to calculate the
emergent spectra for $\beta=0.8$ and $\beta=0.5$. The emergent
spectrum for $\beta=0.8$ is plotted in Figure \ref{disk-beta} with a
red-solid line, and the emergent spectrum for $\beta=0.5$ is plotted
in Figure \ref{disk-beta} with a black-solid line. The dashed lines
are the emissions from the truncated accretion disk. The bolometric
luminosity for $\beta=0.8$ is $ L_{\rm bol} = 9.6\times 10^{42} \
\rm erg \ s^{-1}$, which corresponds to $7.6\times 10^{-4} L_{\rm
Edd}$. The radiative efficiency $\eta$ (defined as $\eta$=$L_{\rm
bol}/\dot M c^2$) of the accretion flow for $\beta=0.8$ is $\eta$ =
$7.7\times 10^{-3}$, which is much lower than the radiative
efficiency $\eta \approx 0.1$ predicted by the standard accretion
disk extending down to the ISCO of a non-rotating black hole. The
bolometric luminosity for $\beta=0.5$ is $ L_{\rm bol} = 3.1\times
10^{43} \ \rm erg \ s^{-1}$, which corresponds to $2.5\times 10^{-3}
L_{\rm Edd}$. The radiative efficiency of the accretion flow for
$\beta=0.5$ is $\eta$ = $0.025$. For comparison, we also plot the
emergent spectrum for $\dot m=0.01$ with the standard accretion disk
extending down to the ISCO of a non-rotating black hole (the
blue-solid line in Figure \ref{disk-beta}). The maximum temperature
of the accretion disk is $\sim 4.2\times 10^4 \rm K$, which has a
emission peaking at $\sim 870.5 \rm \AA$ compared to the peak
emission at $\sim 2075 \rm \AA$ for $\beta=0.5$ and at $\sim 1.1 \mu
m$ for $\beta=0.8$.

It has been  demonstrated that the emergent spectra predicted by the
disk evaporation model can be strongly affected by the magnetic
parameter $\beta$. The effect of $\beta$ to the emergent spectrum is
mainly from the emission of the truncated accretion disk. From
Equation (\ref{tr}), an increase of $\beta$ from 0.5 to 0.8 will
result in  a $(0.8/0.5)^{4.61} \approx 8.7$ times increase of the
truncation radius of the accretion disk. Because the luminosity of
the truncated accretion disk $L_{\rm disk} \propto r_{\rm tr}^{-1}$,
the truncated accretion disk luminosity will decrease by a factor of
8.7.  Meanwhile, because the maximum effective temperature of the
truncated accretion disk $T_{\rm eff, max} \propto r_{\rm
tr}^{-3/4}$, an increase of $\beta$ from 0.5 to 0.8 will result in
the peak emission of the truncated accretion disk shifting towards a
lower frequency by a factor of $\sim 5$ times, e.g., taking $ M=10^8
M_{\odot}$, assuming viscosity parameter $\alpha=0.3$, for $\dot
m=0.01$,  the peak emission of the truncated accretion disk is at
$\sim 2075 \rm \AA$ for $\beta=0.5$ and  at $\sim 1.1 \mu m $ for
$\beta=0.8$. For the inner ADAF, a change of $\beta$ mainly affects
the radio emission, while it has only very little effect on the
optical/UV and X-ray emission. The luminosity of ADAF $L_{\rm ADAF}
\propto \beta$, so a change of $\beta$ from 0.5 to 0.8 will also
imply a little change in the luminosity of the ADAF (Taam et al.
2012; Mahadevan 1997).

\begin{figure*}
\includegraphics[width=85mm,height=70mm,angle=0.0]{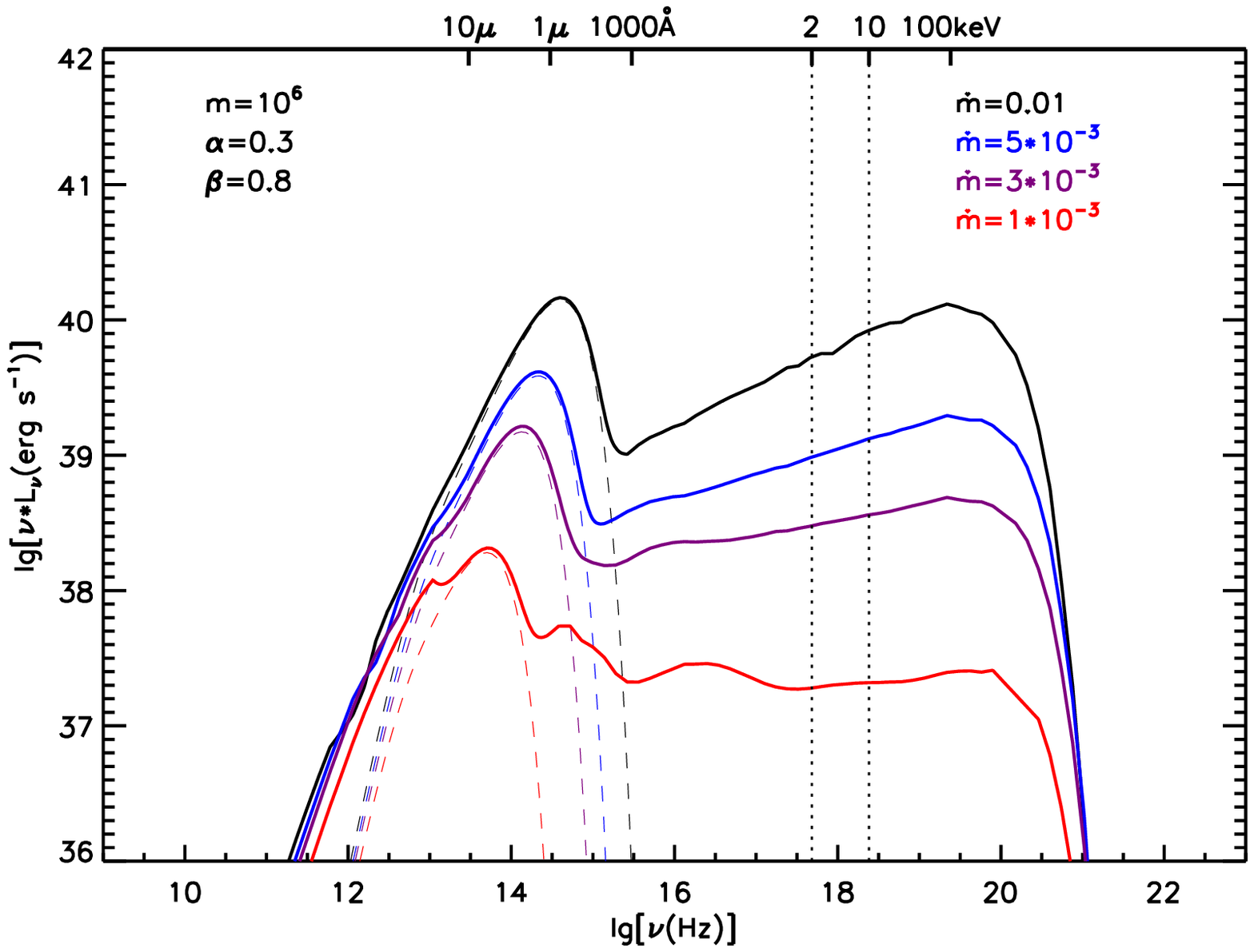}
\includegraphics[width=85mm,height=70mm,angle=0.0]{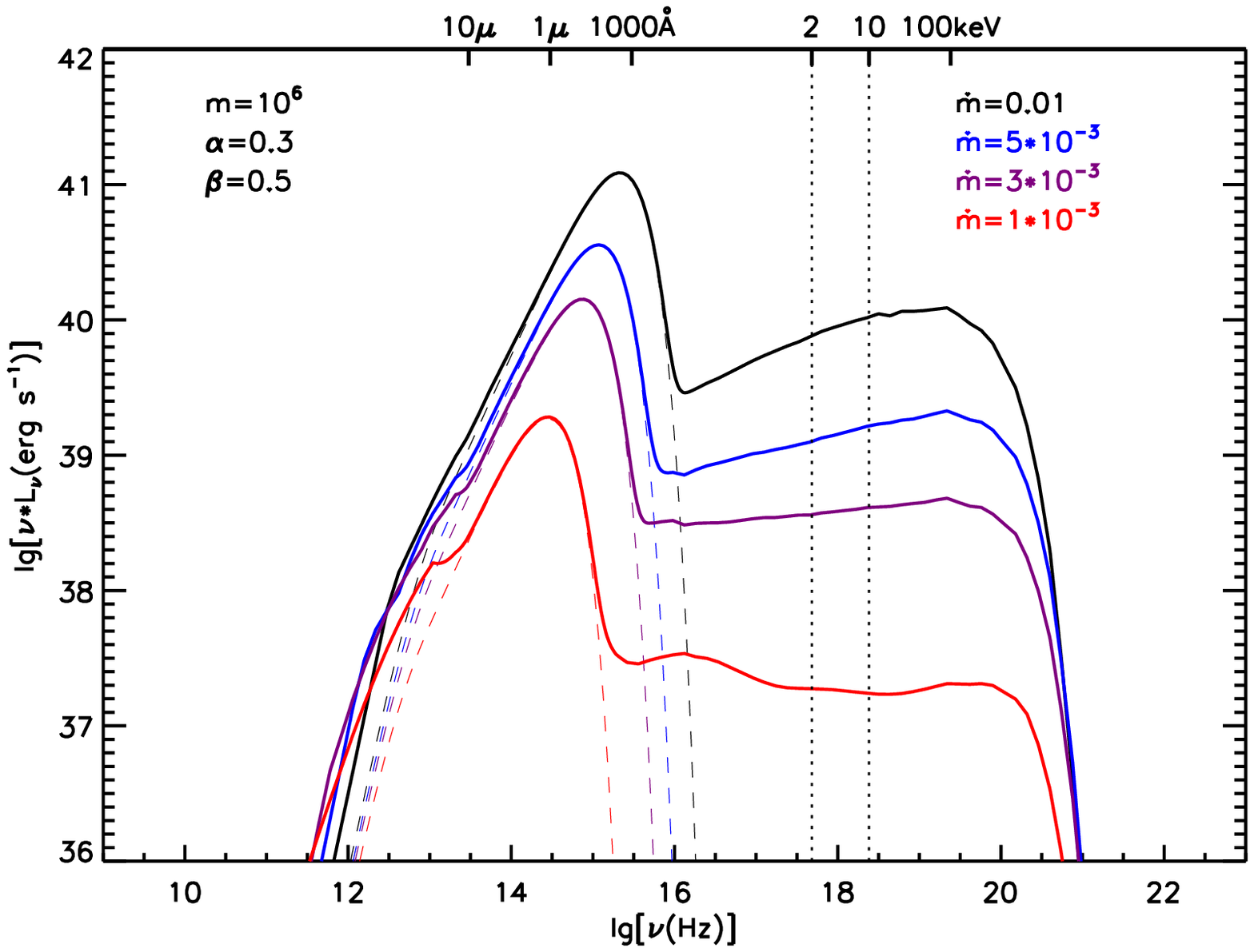}
\caption{\label{disk-adaf06}}Emergent spectra of an inner ADAF and
an outer truncated accretion disk around a black hole predicted by
the disk evaporation model with $M=10^{6} M_{\odot}$ assuming
$\alpha=0.3$. Left panel: $\beta=0.8$ is adopted. From bottom to
top, the solid lines are the combined emergent spectra from an inner
ADAF plus an outer truncated accretion disk for  $\dot m= 10^{-3}$,
$3\times 10^{-3}$, $5\times 10^{-3}$ and $0.01$ respectively. The
dashed lines are the emergent spectra from the truncated accretion
disk. Right panel: $\beta=0.5$ is adopted, and the meaning of the
line style is same with the left panel.
\end{figure*}

\begin{figure*}
\includegraphics[width=85mm,height=70mm,angle=0.0]{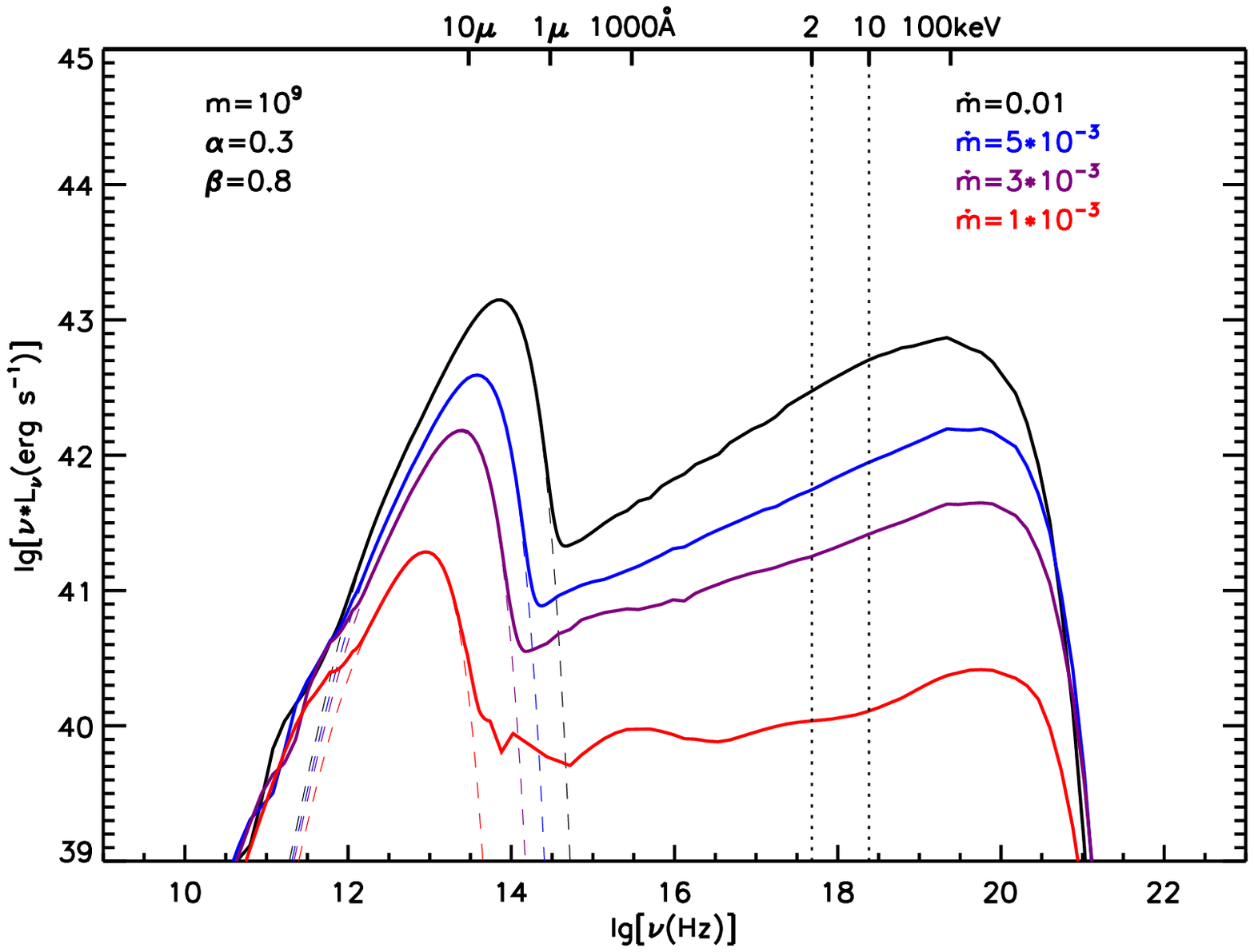}
\includegraphics[width=85mm,height=70mm,angle=0.0]{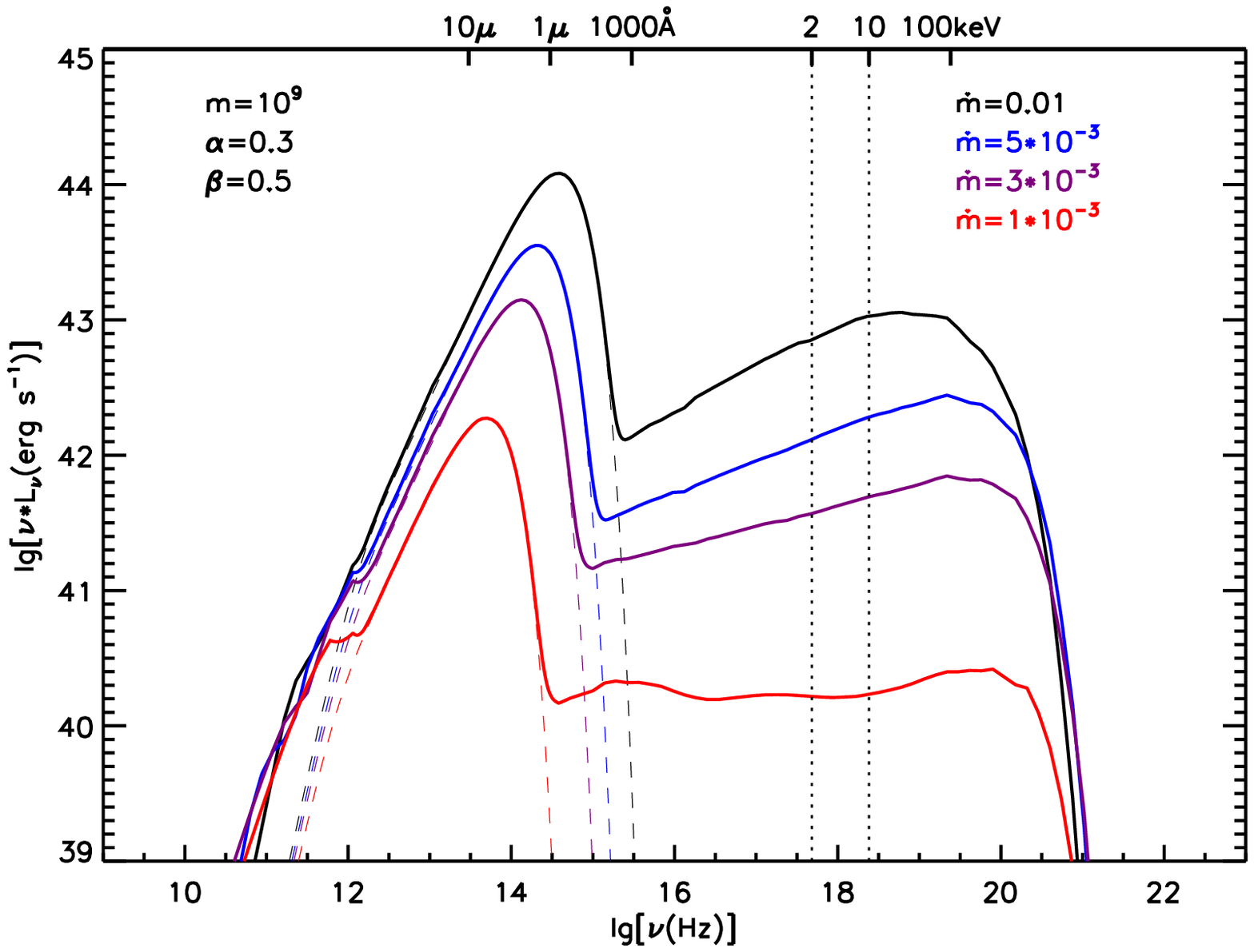}
\caption{\label{disk-adaf09}} Emergent spectra of an inner ADAF and
an outer truncated accretion disk around a black hole predicted by
the disk-evaporation model with $M=10^{9} M_{\odot}$ assuming
$\alpha=0.3$. Left panel: $\beta=0.8$ is adopted. From bottom to
top, the solid lines are the combined emergent spectra from an inner
ADAF plus an outer truncated accretion disk for  $\dot m= 10^{-3}$,
$3\times 10^{-3}$, $5\times 10^{-3}$ and $0.01$ respectively. The
dashed lines are the emergent spectra from the truncated accretion
disk. Right panel: $\beta=0.5$ is adopted, and the meaning of the
line style is same with the left panel.
\end{figure*}

In order to show the  effect of the black hole mass on the emergent
spectra, we plot the emergent spectra for different black hole
masses as comparisons. The emergent spectra for $M=10^6 M_{\odot}$
with mass accretion rates $\dot m=0.01$, $0.005$, $0.003$, $0.001$
are plotted in Figure \ref{disk-adaf06} for the left panel with
$\beta=0.8$, and for the right panel with $\beta=0.5$. The emergent
spectra for $M=10^9 M_{\odot}$ with mass accretion rates $\dot
m=0.01$, $0.005$, $0.003$, $0.001$ are plotted in Figure
\ref{disk-adaf09} for the left panel with $\beta=0.8$, and for the
right panel with $\beta=0.5$.

\section{Comparison with observations---Bolometric Correction $\kappa_{\rm 2-10 keV}$}
Generally, the bolometric luminosity of AGNs is estimated by
multiplying a suitable bolometric correction at a given band. So far
the most secure measurements to the bolometric luminosity is from
X-ray observations. The X-ray bolometric correction $\kappa_{\rm
2-10 keV}$  is determined from the mean energy distribution
calculated from 47 luminous, mostly luminous quasars (Elvis et al.
1994), in which $\kappa_{\rm 2-10 keV} \approx 30$. However, since
the SED of AGNs can change much with mass accretion rate, it is not
a good approximation to correct the bolometric luminosity with a
single correction factor. A correlation between  $\kappa_{\rm 2-10
keV}$ and Eddington ratio $\lambda$ is found by Vasudevan \& Fabian
(2007; 2009), in which $\kappa_{\rm 2-10 keV}$ is  calculated for a
sample with simultaneous X-ray/Optical-UV observations. It is found
that $\kappa_{\rm 2-10 keV} \approx 15-30$ for $\lambda \lesssim
0.1$, $\kappa_{\rm 2-10 keV} \approx 20-70$ for $ 0.1 \lesssim
\lambda \lesssim 0.2$, and $\kappa_{\rm 2-10 keV} \approx 70-150$
for $\lambda \gtrsim  0.2$ respectively. By compiling the SEDs of a
small  LLAGN sample, Ho (1999b) found that the median bolometric
correction is $\kappa_{\rm 2-10 keV} \approx 8$. A more extensive
data set suggested a value larger by a factor of 2 (Ho 2000; 2009),
having a median value $\kappa_{\rm 2-10 keV} \approx 15.8$. Because
LLAGNs tend to be "X-ray-loud", they have smaller values of
$\kappa_{\rm 2-10 keV}$ compared with the luminous sources. This is
consistent with the results of vasudevan \& Fabian (2007; 2009) and
Elvis et al. (1994).

\begin{figure*}
\includegraphics[width=85mm,height=70mm,angle=0.0]{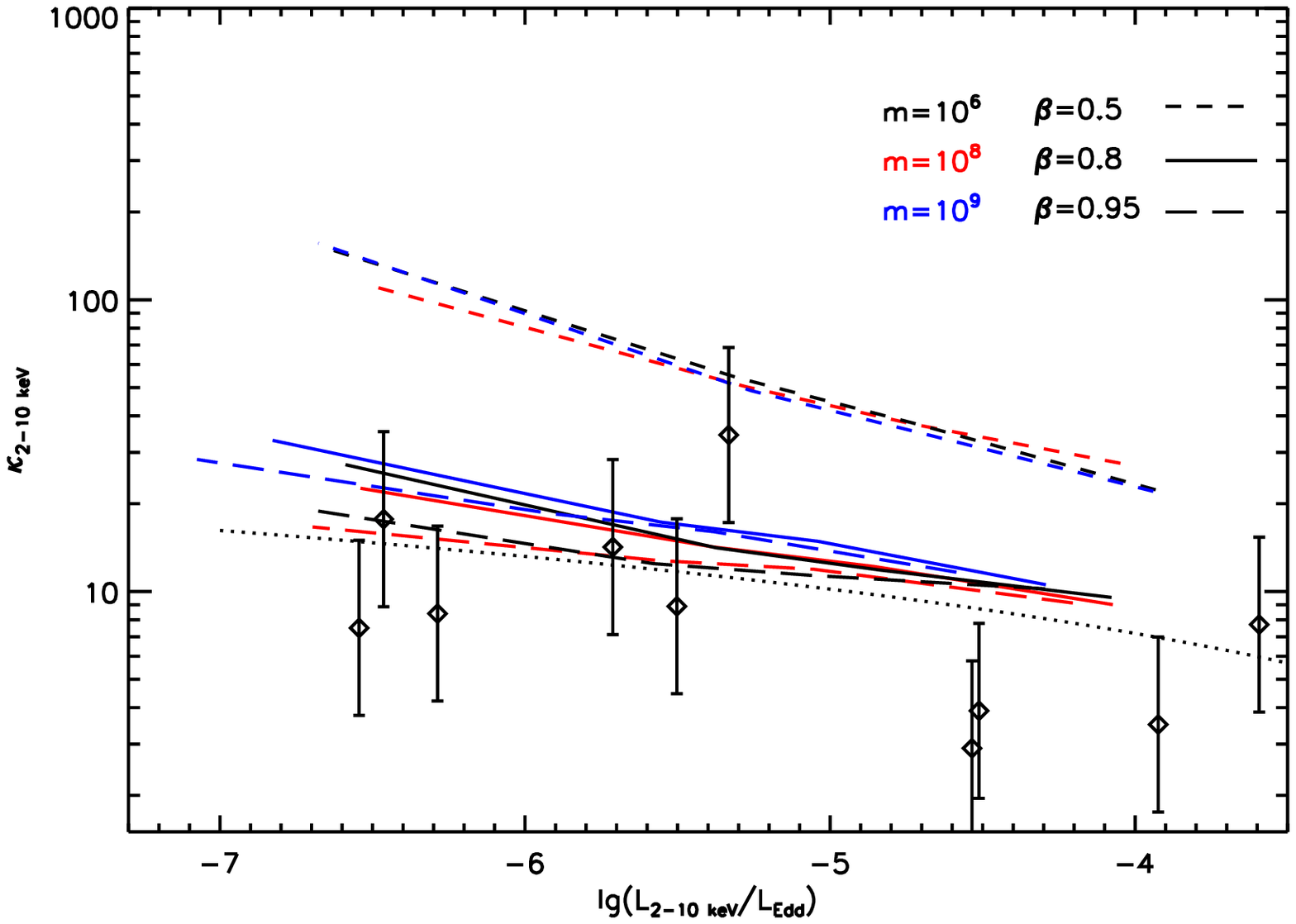}
\caption{\label{kappa}} Bolometric correction $\kappa_{\rm 2-10keV}$
as a function of $L_{\rm 2-10keV}/L_{\rm Edd}$. The short-dashed
line is for $\beta=0.5$, the solid line is for $\beta=0.8$ and the
long-dashed line is for $\beta=0.95$. The black line is for $M=10^6
M_{\odot}$, red line is  $M=10^8 M_{\odot}$ and the blue line is for
$M=10^9 M_{\odot}$. In all the calculation, $\alpha=0.3$ is adopted.
The sign $\diamondsuit$ is the observed data, including NGC 1097,
NGC 3031, NGC 4203, NGC 4261, NGC 4374, NGC 4450, NGC 4486, NGC
4579, NGC 4594 and NGC 6251 with  2-10 keV luminosity $L_{\rm 2-10
keV}$ measurement and bolometric luminosity measurement from Ho
(2009).
\end{figure*}

We collect a sample composed of 10 LLAGNs, including NGC 1097, NGC
3031, NGC 4203, NGC 4261, NGC 4374, NGC 4450, NGC 4486, NGC 4579,
NGC 4594 and NGC 6251 with  2-10 keV luminosity $L_{\rm 2-10 keV}$
measurement and bolometric luminosity measurement from Ho (2009).
The bolometric luminosity is obtained by integrating the
interpolated SEDs shown in Ho (1999b) and Ho et al. (2000). The
black hole masses are also collected by Ho (1999b) and Ho (2000).
The bolometric correction $\kappa_{\rm 2-10 keV}$ as a function of
$L_{\rm 2-10keV}/$ $L_{\rm Edd}$ is plotted with the sign
$\diamondsuit$ in Figure \ref{kappa}. Ho (2009) conservatively
estimated that the errors of $ L_{\rm bol}/L_{\rm 2-10keV} $ of the
sources in the sample should be within 0.3 dex. The best-fitting
linear regression for the correlation between $\kappa_{\rm 2-10
keV}$ and $L_{\rm 2-10keV}/$ $L_{\rm Edd}$ is as follows:
\begin{eqnarray}\label{fit}
\kappa_{\rm 2-10keV}=-4.8-3.0 \times {\rm log_{10}} (L_{\rm
2-10keV}/L_{\rm Edd}),
\end{eqnarray}
plotted as a dotted-line in Figure \ref{kappa}.

In order to compare with observations, we calculate the emergent
spectra predicted by the disk evaporation model. As shown in section
3.1, magnetic parameter $\beta$ has significant effects on the
emergent spectrum, we calculate the emergent spectra for
$\beta=0.8$, $\beta=0.95$, and $\beta=0.5$ for comparisons. By
integrating the emergent spectrum, we calculate the 2-10 keV
luminosity $L_{\rm 2-10keV}$ and the bolometric luminosity $L_{\rm
bol}$, with which we can calculate $\kappa_{\rm 2-10 keV}$ and
$L_{\rm 2-10keV}/$ $L_{\rm Edd}$.

Fixing $\beta=0.8$, taking $ M=10^8 M_{\odot}$ and assuming
$\alpha=0.3$, the ratio of 2-10keV luminosity $L_{\rm 2-10keV}$ to
Eddington luminosity $ L_{\rm Edd}$ is $L_{\rm 2-10keV}/L_{\rm Edd}$
= $8.45\times 10^{-5}$, $1.40 \times 10^{-5}$, $4.13\times 10^{-6}$,
$2.90 \times 10^{-7}$, and the bolometric correction is $\kappa_{\rm
2-10 keV}= 9.0, 12.2, 14.2, 22.6$ for $\dot m= 0.01. 0.005, 0.003,
0.001$ respectively. $\kappa_{\rm 2-10 keV}$ as a function of
$L_{\rm 2-10keV}/L_{\rm Edd}$ is plotted with a red-solid line in
Figure \ref{kappa}. It can be seen that there is an anti-correlation
between $L_{\rm 2-10 \rm keV}/L_{\rm Edd}$ and $\kappa_{\rm 2-10
keV}$.  In order to check the effect of the black hole mass on
$L_{\rm 2-10keV}/L_{\rm bol}$ - $\kappa_{\rm 2-10 keV}$ correlation,
we take different black hole masses for comparisons. For $ M=10^6
M_{\odot}$, $\kappa_{\rm 2-10 keV}$ as a function of $L_{\rm
2-10keV}/L_{\rm bol}$ is plotted with a black-solid line in Figure
\ref{kappa}. For $ M=10^9 M_{\odot}$, $\kappa_{\rm 2-10 keV}$ as a
function of $L_{\rm 2-10keV}/L_{\rm bol}$ is plotted with a
blue-solid line in Figure \ref{kappa}. It can be seen that the
effect of the black hole mass on $L_{\rm 2-10keV}/L_{\rm Edd}$ -
$\kappa_{\rm 2-10 keV}$ correlation is very weak.  As an extreme
example, we also plot $L_{\rm 2-10keV}/L_{\rm Edd}$ - $\kappa_{\rm
2-10 keV}$ correlation for $\beta=0.95$ in Figure \ref{kappa}. The
red long-dashed, black long-dashed, blue long-dashed lines are for
$M=10^8 M_{\odot}$, $M=10^6 M_{\odot}$ and $M=10^9 M_{\odot}$
respectively.

Fixing $\beta=0.5$, taking $M=10^8 M_{\odot}$ and assuming
$\alpha=0.3$, the ratio of 2-10keV luminosity $L_{\rm 2-10keV}$ to
Eddington luminosity $ L_{\rm Edd}$ is $L_{\rm 2-10keV}/L_{\rm Edd}$
= $8.94\times 10^{-5}$, $1.87 \times 10^{-5}$, $5.32 \times
10^{-6}$, $3.04 \times 10^{-7}$, and the bolometric correction is
$\kappa_{\rm 2-10 keV}= 27.5, 37.5, 50.3, 112.6$ respectively.
$\kappa_{\rm 2-10 keV}$ as a function of $L_{\rm 2-10keV}/L_{\rm
Edd}$ is plotted with a red short-dashed line in Figure \ref{kappa}.
The bolometric correction $\kappa_{\rm 2-10 keV}$ as a function of
$L_{\rm 2-10keV}/L_{\rm Edd}$ for $M=10^6 M_{\odot}$ is plotted with
a black short-dashed line in Figure \ref{kappa}, and for $M=10^9
M_{\odot}$ is plotted with a blue short-dashed line in Figure
\ref{kappa}.

From Figure \ref{kappa}, it is clear that the bolometric correction
strongly depends on $\beta$. $L_{\rm 2-10keV}$ $/L_{\rm bol}$ -
$\kappa_{\rm 2-10 keV}$ correlation predicted by $\beta=0.8$ or
$\beta=0.95$ can well explain the observations, while the prediction
by $\beta=0.5$ severely deviates from the observations. Although our
model for larger $\beta$ can roughly interpret the $L_{\rm 2-10keV}$
$/L_{\rm bol}$ - $\kappa_{\rm 2-10 keV}$ correlation, it is still a
few sources lay below the model line. This may be because, in this
work, for simplicity, we assume that, outside the truncation radius,
the accretion flow exists in the form of a pure standard accretion
disk. However, according to the disk evaporation model, outside the
truncation radius, the matter should be in the form of
"disk+corona", but not the pure stand accretion disk. The emission
from the corona will result in an increase of the X-ray emission,
which will make the bolometric correction decrease. Consequently,
the model-predicted $\kappa_{\rm 2-10 keV}$ will systematically
shift downward. We also need to keep in mind that, due to the inner
ADAF dominating the X-ray emission, the emission contribution from
the outer corona to the bolometric correction will be very small, so
still the data support a bigger value of $\beta=0.8$ or
$\beta=0.95$.

Theoretically, it would be useful to understand how the magnetic
parameter is determined and how it is constrained. The value of
$\beta$ can be approximately known from the numerical simulations
(e.g., Balbus \& Hawley 1998). Detailed MHD numerical simulations
for the formation of a magnetized corona have shown that a strongly
magnetized corona can form above an initially weakly magnetized disk
(e.g., Miller \& Stone 2000; Machida et al. 2000; Hawley \& Balbus
2002). However, in the disk evaporation model, we consider a slab
corona in a large vertical extent, in which the corona is vertically
stratified, in contrast to an isothermal torus as from MHD
simulations. Furthermore, thermal conduction in vertical direction
is taken into account, which will make an efficient mass evaporation
from the disk to the corona, resulting in a higher mass density than
in the coronal envelope seen in MHD simulations. More importantly,
in the disk evaporation model, the electrons in the corona are
assumed to be heated only by a transfer of energy from the ions to
electrons via Coulomb collisions, which is reasonable for the
accretion with a lower mass accretion rate (Narayan 1995b). Coulomb
heating is the dominated heating mechanism for the electrons only if
the magnetic field is strongly sub-equipartition (i.e., ratio of gas
pressure to magnetic pressure $>$ 10) (Malzac \& Belmont 2009). The
spectral modeling of Sgr $A^{*}$ showed that an additional direct
heating to electrons is required, which is probably produced by
magnetohydrodynamics turbulence, magnetic reconnection and weak
shocks (Yuan et al. 2003). However, the magnetic heating mechanism
to the electrons is very unclear, consequently still a weak magnetic
field, i.e., a high value of $\beta$ is assured in the case of a
lower mass accretion rate within the framework of disk evaporation
model.


\section{Conclusion}

Based on the prediction by Taam et al. (2012) of a truncation radius
of the accretion disk on the  mass accretion rate by including the
magnetic field,  we have calculated the emergent spectrum of an
inner ADAF + an outer truncated accretion disk around a
super-massive black hole. It is found that the disk evaporation
model can roughly reproduce the observed anti-correlation between
hard X-ray index $\Gamma_{\rm 2-10 \rm keV}$ and Eddington ratio
$\lambda$ for $\lambda \lesssim 10^{-3}$ in LLAGNs. As shown by Taam
et al. (2012), the truncation radius of the accretion disk is
sensitive to the magnetic parameter $\beta$, which consequently
affects the emergent spectrum. Our calculations show that the
equipartition of gas pressure to magnetic pressure, i.e.,
$\beta=0.5$, failed to explain the observed anti-correlation between
$L_{\rm 2-10 keV}/L_{\rm Edd}$ and the bolometric correction
$\kappa_{\rm 2-10 keV}$. The resulted spectra for larger value
$\beta=0.8$ or $\beta=0.95$ can better explain the observed $L_{\rm
2-10 keV}/L_{\rm Edd}$-$\kappa_{\rm 2-10 keV}$ correlation. It is
argued that in the disk evaporation model, the electrons in the
corona are assumed to be heated only by a transfer of energy from
the ions to electrons via Coulomb collisions, which is reasonable
for the accretion with a lower mass accretion rate. Coulomb heating
is the dominated heating mechanism for the electrons only if the
magnetic field is strongly sub-equipartition, which is roughly
consistent with observations.







We thank the very useful discussions with Prof. R. E. Taam from
Northwestern University. This work is supported by the National
Natural Science Foundation of China (grants 11033007 and 11173029),
by the National Basic Re-search Program of China-973 Program
2009CB824800.


\end{document}